\documentclass[review]{elsarticle}
\usepackage{adjustbox,amsfonts,amsmath,caption,dirtytalk,graphicx,pgfplots,subfigure,tikz,lineno,hyperref,xcolor,bm,amssymb,mathtools}
\usetikzlibrary{positioning,fit,shapes.geometric,backgrounds,plotmarks}
\usetikzlibrary{external}
\newcommand{\cblue}{\textcolor{black}}
\graphicspath{{Figures/}}
\pgfplotsset{compat=1.17}
\journal{}

\begin{document}

\begin{frontmatter}

\title{Denoising of discrete-time chaotic signals using echo state networks}
\author{André L. O. Duarte}
\author{Marcio Eisencraft}
\address{Escola Politécnica, University of São Paulo, São Paulo, Brazil}
\begin{abstract}
\cblue{Noise reduction is a relevant topic in the application of chaotic signals in communication systems\cblue{,} in modeling bi\-o\-med\-i\-cal signals \cblue{or in time series forecasting}.
In this paper an echo state network (ESN) is employed to denoise a discrete-time chaotic signal corrupted by additive white Gaussian noise. The choice of applying ESNs in this context is motivated by their successful exploitation for the separation and prediction of continuous-time chaotic signals. 
Our results show that the processing gain of the ESN is higher than the one obtained using a Wiener filter for chaotic signals generated by a skew-tent map. Since the power spectral density of the orbits in this map is well known, it was possible to analyze how the processing gain of the ESN in the denoising process varies according to the spectral characteristics of the chaotic signals. }
\end{abstract}


\end{frontmatter}

\section{Introduction\label{sec:introd}}
Many different techniques have been proposed in an effort to solve the noise reduction problem \cite{Vaseghi2009}. In recent years, there has been an increased interest in machine learning techniques due to their great adaptability, despite their rather simple design principles. Therefore, it is natural to investigate their performance for noise mitigation \cblue{compared to other well-established techniques such as wavelet filtering \cite{Han2009,Gao2010}}. 

In particular, Echo State Networks (ESNs) emerged in the early 2000s in an effort to reduce the problems faced when designing Recurrent Neural Networks (RNNs): (i) the training of RNNs is cumbersome when using classical methods based on gradient descent, such as backpropagation; (ii) the lack of convergence guarantee; and (iii) the high computational effort required due to the large number of parameters to optimize \cite{Jaeger}.

The distinguishing feature of ESNs is that only the weights in the output layer are optimized during training \cite{Jaeger}.  The remaining weights do not change during training, minimizing the number of parameters to optimize \cite{Jaeger}. ESNs are widely used due to their simplicity and low computational cost in many applications related to dynamical systems, such as the separation \cite{Hunt} and prediction \cite{Xu2019} of chaotic signals. \textcolor{black}{The computation time for a given performance is much lower than that of a standard RNN, reaching over 99\% in some cases \cite{Shahi2022}.}

\textcolor{black}{An investigation of the use of ESN for noise reduction in sinusoidal waves with random phases and time-varying envelopes was carried out by \cite{OliveiraJunior2022}. At \cite{Sun2021}, ESNs were used to denoise electroencephalogram signals disturbed by physiological signals and electrode interference. In \cite{Han2021}, an ESN is used together with a Kalman filter to equalize a chaotic signal in a nonlinear channel.} 

In the present work, \textcolor{black}{we consider discrete-time} chaotic signals disturbed by additive white Gaussian noise (AWGN). The aperiodic behavior, together with the sensitive dependence on initial conditions of chaotic signals \cite{Alligood} adds to the difficulty of the denoising task. \textcolor{black}{To demonstrate that the ESN can be useful in solving the denoising problem, we compare the processing gain of ESN with that of the Wiener Filter (WF), as it is the optimal linear filtering technique.}

\cblue{It is important to note that in this paper the chaotic signals are generated by a discrete-time system, in contrast to most of the literature, which generally considers standard continuous-time dynamical systems, such as Lorenz or Rossler systems \cite{Han2007,Lou2022}. This is relevant not only because digital signal processors and simulation software inherently use discrete-time signals, but also because many processes, such as digital currencies \cite{Altan2019} and crude oil time series \cite{Karasu2022}, generate chaotic signals in discrete time. These signals can also represent equivalent low-pass discrete-time signals in chaos-based communication systems \cite{Kennedy2000a}.}

In particular, the skew tent map \cite{Eisencraft2010} is used to obtain the chaotic signals here. The power spectral density (PSD) of its orbits is determined by its single parameter in a simple deterministic way \cite{Eisencraft2010}. The rationale for considering this map here is thus justified, because knowing the PSD in the context of noise reduction is useful for performing an analysis of how orbits with different PSDs affect the denoising task.

\begin{figure}[htb]
    \centering
    \includegraphics[]{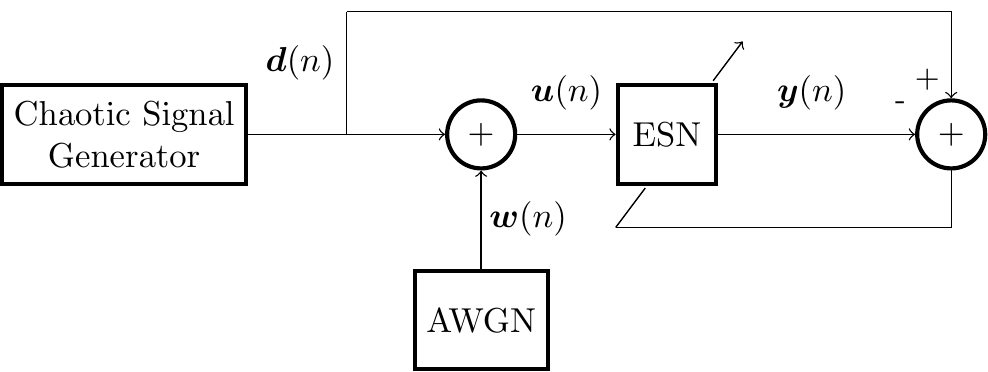}
    \caption{Problem formulation: noise reduction of a chaotic signal using ESN.}
    \label{fig:blockDiagram}
\end{figure}

Figure \ref{fig:blockDiagram} shows a block diagram of the problem at hand.
After the training period, the ESN output $\bm{y}(n)$ is expected to approximate the desired noiseless chaotic signal $\bm{d}(n)$ from the corrupted input signal $\bm{u}(n)=\bm{d}(n)+\bm w(n)$, where $\bm{w}(n)$ is AWGN.

The remainder of this short communication is divided into four sections. Section \ref{sec:chaoticSignal} gives a brief description of the chaotic signal generator considered here. Section \ref{sec:echoStateNetworks} discusses the operation of the ESN.  Section \ref{sec:results} presents and discusses the results obtained. Finally, our main conclusions are summarized in Section \ref{sec:conclusions}.
\section{Considered Chaotic Signals}
\label{sec:chaoticSignal}
\begin{figure}[htb]
    \centering
    \includegraphics[width=\linewidth,trim={0 0.8cm 0 0}]{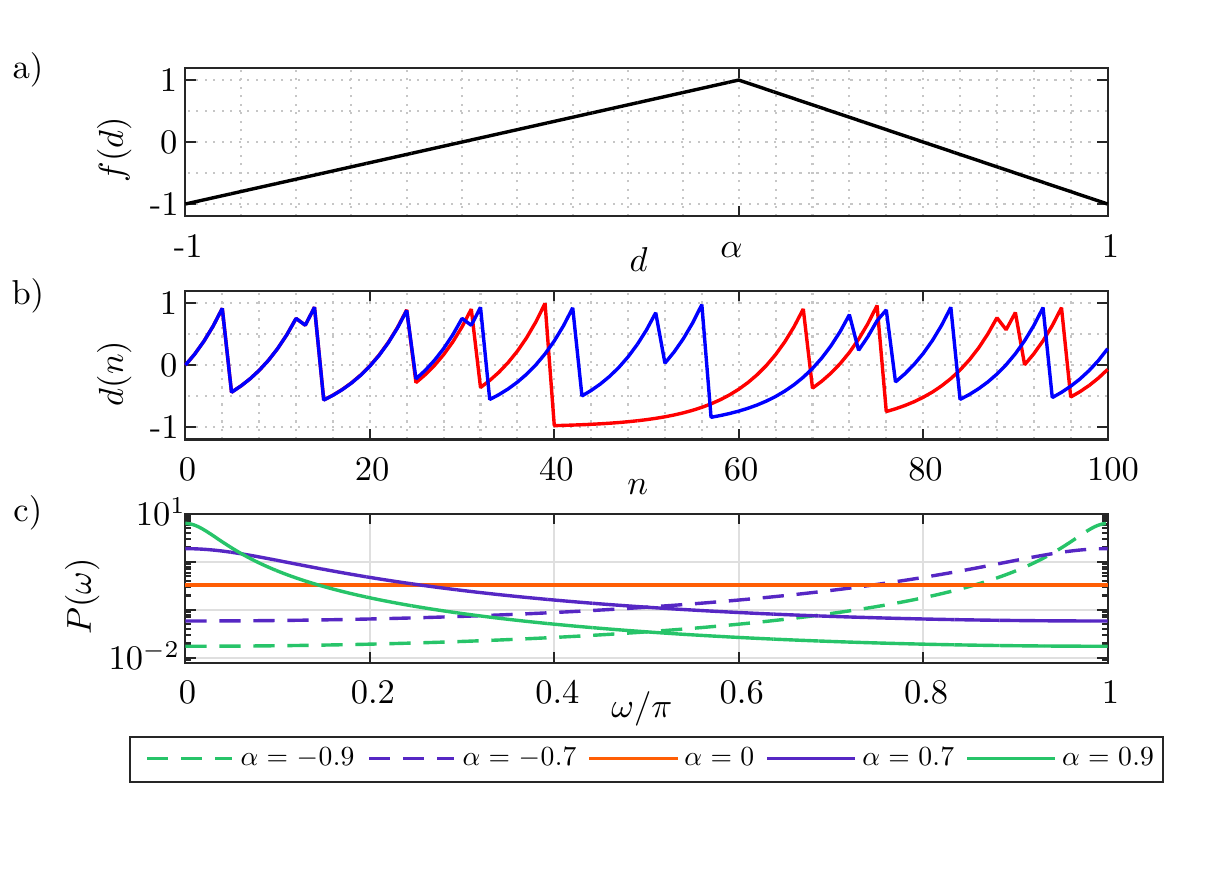}
    \caption{a) The skew tent map \eqref{eq:skewTentMap}; b) two orbits for  $\alpha=0.7$ with initial conditions $d_0=0$ (red line) and $d_0=10^{-6}$ (blue line) showing SDIC; c) PSD for different values of $\alpha$.}
    \label{fig:chaoticSignalsFigure}
\end{figure}

Chaotic signals are bounded, aperiodic, and have a sensitive dependence on initial conditions (SDIC) \cite{Alligood}.
The one-dimensional discrete-time chaotic signal $d(n)$ considered in this paper is obtained from the skew tent map defined by \cite{Eisencraft2010}
\begin{equation}
    d(n+1) = f(d(n)) = \begin{cases}
    \frac{1-\alpha}{1+\alpha}+\frac{2}{1+\alpha}d(n), & -1 < d(n)<\alpha \\
          \frac{1+\alpha}{1-\alpha}-\frac{2}{1-\alpha}d(n),& \alpha \leq d(n) < 1
    \end{cases},
    \label{eq:skewTentMap}
\end{equation}
with initial condition $d(0)\triangleq d_0\in (-1, 1)$ and fixed parameter $\alpha \in (-1, 1)$, the $x$-coordinate of the peak of the tent. Figure \ref{fig:chaoticSignalsFigure}(a) shows a plot of $f(\cdot)$ and Figure \ref{fig:chaoticSignalsFigure}(b) depicts two of its orbits, with slightly different initial conditions, to illustrate the SDIC property.

In \cite{Eisencraft2010} it was shown that the PSD of the orbits of $f(\cdot)$ is given by 
\begin{equation}
    P(\omega) = \frac{1-\alpha^2}{3(1+\alpha^2-2\alpha\cos{\omega})}.
    \label{eq:skewTentMapPSD}
\end{equation}
As shown in Figure \ref{fig:chaoticSignalsFigure}(c), $P(\omega)$ is white for $\alpha=0$. \textcolor{black}{The essential bandwidth \cite{Eisencraft2010} of the chaotic signals decreases with $|\alpha|$, and the PSD is concentrated at higher frequencies ($\alpha<0$) or lower frequencies ($\alpha>0$).} In other words, the further away from $\alpha=0$, the more the spectral characteristics of the chaotic signal generated will differ from a white noise signal. This direct relationship between the $\alpha$ parameter and the PSD is what led us to choose this map. It allows us to analyze, in Section \ref{sec:results}, the influence of the PSD of the chaotic signal on the performance of the ESN in the noise reduction task.

\section{Echo State Networks}
\label{sec:echoStateNetworks}
\begin{figure}[htb]
    \centering
    \includegraphics[width=\linewidth]{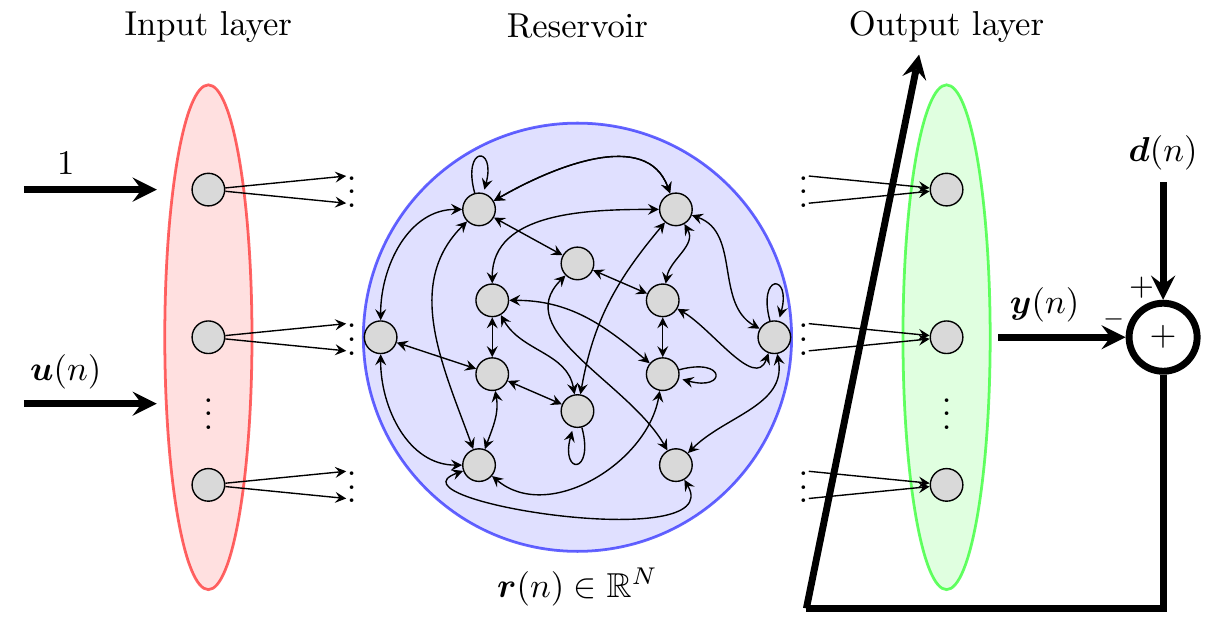}
    \caption{ESN architecture, indicating the signals $\bm{u}(n)$, $\bm{y}(n)$, $\bm{d}(n)$ and the internal state vector  $\bm{r}(n)$.}
    \label{fig:ESNarchitecture}
\end{figure}

The purpose of an ESN as shown in Figure \ref{fig:ESNarchitecture} is to use an input signal $\bm{u}(n)\in \mathbb{R}^{N_u}$ to approximate a target signal $\bm{d}(n)\in \mathbb{R}^{N_d}$ after a training period. It consists of (i) an input layer, (ii) the so-called reservoir, and (iii) an output layer. Each of these parts is made up of information processing nodes connected by links, through which information is exchanged \cite{Jaeger,Lukosevicius2012A}.

The role of the input layer is to preprocess the input signal in order to control the amount of nonlinearity of the ESN \cite{Lukosevicius2012A}. It computes $\bm{W}^{\text{in}}\begin{bsmallmatrix}1 & \bm{u}(n)\end{bsmallmatrix}^T$, where $\bm{W}^{\text{in}}$ is $N \times N_u+1$.
So there are $N_u+1$ entry nodes, one for each dimension of $\bm{u}(n)$ and an extra one for the bias.
The matrix $\bm{W}^\text{in}$ is called the \emph{input matrix} and its entries are samples from uniform distributions. The entries in the first column are drawn from a uniform distribution in $[-p, p]$, while the remaining entries follow a uniform distribution in $[-q, q]$. 

The reservoir consists of $N$ nodes connected to each other. The weight of each link is determined randomly and are the entries of the \emph{internal matrix} $\bm{W}_{N \times N}$. They are obtained as follows: first, the entries of an auxiliary matrix $\bm{W}^\text{aux}_{N \times N}$ are drawn from a uniform distribution in $[-1, 1]$. Then its spectral radius $\rho$ is determined. Finally, we compute $\bm{W}=\lambda\left(\bm{W}^\text{aux}/\rho\right)$, where $\lambda$ is a parameter. Since $\bm{W}^\text{aux}/\rho$ has unit spectral radius, it follows that $\bm{W}$ has spectral radius $\lambda$. 

Each node $k$, $1\leq k\leq N$, in the reservoir has an internal state $r_k \in \mathbb{R}$, which forms the internal state vector $\bm{r}(n) \in \mathbb{R}^N$. At each time $n+1$, the internal state vector is updated according to the leaky-integrator model equation
\begin{equation}
    \bm{r}(n+1) = \left(1-a\right)\bm{r}(n)+a\tanh\left(\bm{W}^{\text{in}}\begin{bmatrix}
 1 & \bm{u}(n)
 \end{bmatrix}^T+\bm{W}\bm{r}(n)\right),
    \label{eq:stateVectorUpdate}
\end{equation}
with leakage parameter  $a\in[0, 1]$ and initial condition $\bm{r}(-\ell+1)=\bm{0}$. 

After a transient period of $\ell$ time steps, the training period begins. During $L \in \mathbb{N}$ training time steps, the corresponding state vectors are collected in the trajectory matrix $\bm{T}_{N\times L}$, defined by $
\bm{T} = \begin{bsmallmatrix}
 \bm{r}(1) & \bm{r}(2) & \dots & \bm{r}(L)
 \end{bsmallmatrix}.$
which is passed to the output layer.

Let $\bm{D}_{N_d\times L}$ be the matrix of samples of the desired signal available for training, that is, $ \bm{D} = \begin{bsmallmatrix}
     \bm{d}(1) & \bm{d}(2) & \dots & \bm{d}(L)
     \end{bsmallmatrix}.
     \label{eq:desiredMatrix}
$
The optimized weights of the connections from the reservoir to the output are given by $
     \bm{W}^\text{out} = \bm{D}\bm{T}^{+}$,
where $\bm{T}^{+}$ is the Moore-Penrose pseudoinverse of $\bm{T}$. 
After training, the ESN output signal is calculated as a linear combination,
$
\bm{y}(n) = \bm{W}^\text{out}\bm{r}(n)$.    
    
At this point, our description of the ESN is almost complete. Only four parameters remain to be adjusted: (i)  the leakage parameter $a$, (ii) the spectral radius $\lambda$ of $\bm{W}$, and the scalars (iii) $p$ and (iv) $q$ that define the intervals of the uniform distributions used to obtain $\bm{W}^\text{in}$. Our methodology for choosing them is described along with the numerical results in Section \ref{sec:results}.

\section{Numerical simulations}
\label{sec:results}
\begin{figure}[htb]
    \centering
     \includegraphics[width=\linewidth,trim={0 2cm 0 0}]{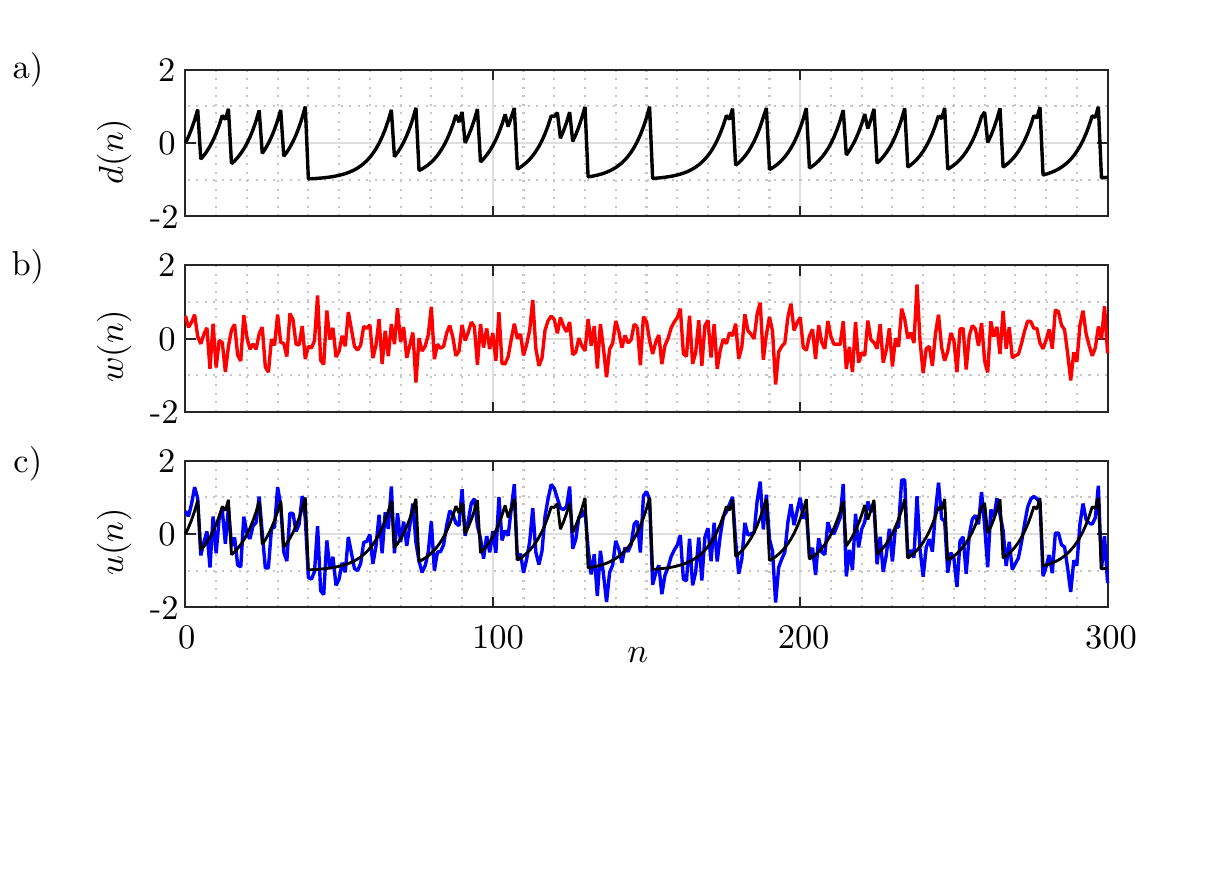}
    \caption{Examples of training signals: a) noiseless chaotic signal $d(n)$ for $\alpha=0.7$ and $d_0=0$; b) AWGN for $\text{SNR}_\text{in} = 2.0$~dB; c) input signal $u(n)$ (blue line) for the ESN; $d(n)$ is also shown for comparison.}
    \label{fig:inputSignal}
\end{figure}

The desired one-dimensional  chaotic signal $d(n)$ is generated by \eqref{eq:skewTentMap} for a given $\alpha$ and initial condition $d_0$. The  corrupted input $u(n)$ of the ESN is then $u(n)=d(n)+w(n)$, 
where the AWGN  $w(n)$ power is determined by a chosen $\text{SNR}_\text{in}$. Figure \ref{fig:inputSignal} shows examples of $d(n)$, $w(n)$ and $u(n)$ for $d_0=0$, $\alpha=0.7$ and $\text{SNR}_\text{in}=2.0$~dB.

\vspace{-.2cm}

Our goal is to train the ESN to reduce the noise component in $u(n)$. This means that after sufficient training, the output signal $y(n)$ of the network should mimic the desired chaotic signal $d(n)$, using only $u(n)$ as input.

The ESN was designed with $N=500$ nodes, a transient of $\ell=200$ samples, and $L=25000$ samples for training. For the estimation of the output SNR ($\text{SNR}_\text{out}$), \color{black}$L_T=10^6$ samples were considered in the ratio
\begin{equation}
    \text{SNR}_\text{out} = \frac{\sum_{n=L+1}^{L_T}y^2(n)}{\sum_{n=L+1}^{L_T}(d(n)-y(n))^2}.
    \label{eq:SNRout}
\end{equation}\normalcolor

\subsection{Selecting the ESN parameters\label{subsec:parameters}}

\color{black}To select the ESN parameters $a$, $\lambda$, $p$ and $q$, we have generated $u(n)$ with an arbitrary $d_0 \in (-1, 1)$, $\alpha = 0.1$ and $\text{SNR}_\text{in} = 2.0$~dB. The selection routine can be summarized in the following steps:
\begin{small}
\begin{enumerate}
    \item Start with $\lambda=\lambda_0 = 0.05$, $p=p_0 = 0$ and $q=q_0 = 0.5$.
    
    \item Keeping $\lambda$, $p$ and $q$ fixed, vary $a$ from $0$ to $1$ in $0.05$ steps. In each case, train and test an ESN to find $a=a_1$ that maximizes $G[\text{dB}] \triangleq\text{SNR }_\text{out}\text{[dB]}-\text{SNR}_\text{in}\text{[dB]}$.

    \item Similarly, $\lambda_1$ is determined by varying $\lambda$ between $0.05$ and $1$ in $0.05$ steps, using $a=a_1$, $p=p_0$ and $q=q_0$.

    \item Next, $p_1$ is obtained by varying $p$ between 0 and 10 in $0.5$ steps, with $a=a_1$, $\lambda=\lambda_1$ and $q=q_0$.

    \item Similarly, $q_1$ is obtained by varying $q$ between 0.5 and 10 in $0.5$ steps, while $a=a_1$, $\lambda=\lambda_1$ and $p=p_1$.

    \item Steps 2-5 are repeated, yielding $a_i$, $\lambda_i$, $p_i$, $q_i$, $i=2,3,\ldots$. The search continues until the selected values for each parameter do not change. They correspond to the selected parameters $a_\star$, $\lambda_\star$, $p_\star$, and $q_\star$.
\end{enumerate}
\end{small}
\normalcolor

\begin{figure}[htb]
    \centering
 \includegraphics[width=\linewidth,trim={0 5mm 0 0}]{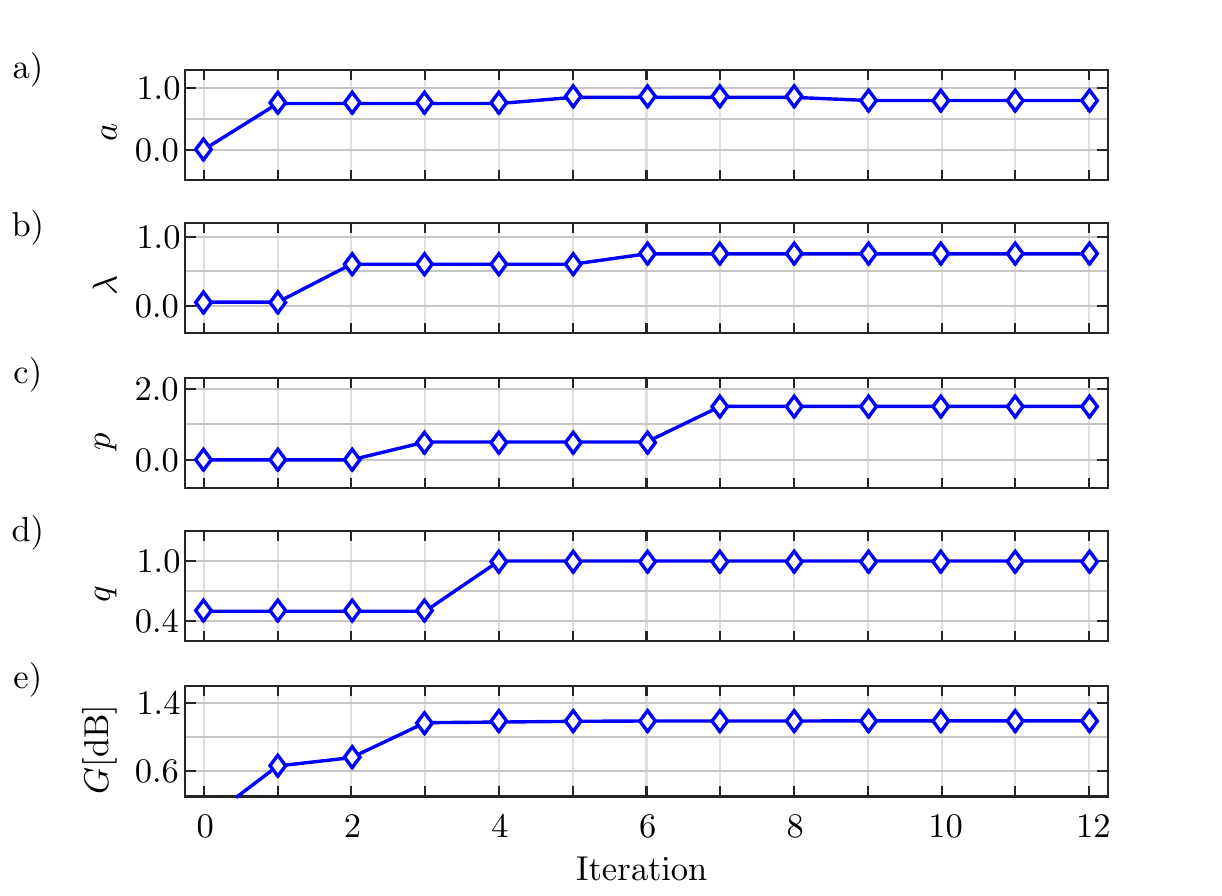}
    \caption{Selection of the ESN parameters. Parameters at each iteration: a) $a$: leakage parameter ; b) $\lambda$: spectral radius of $\bm{W}$; c) $p$ and d) $q$: parameters of uniform distributions used to determine the entries of $\bm{W}^\text{in}$; e) processing gain $G [\text{dB}]$ at each iteration.
    }
    \label{fig:parametersOptimization}
\end{figure}

A typical evolution of the parameters during this procedure is shown in Figure \ref{fig:parametersOptimization}.
It took 12 iterations for all parameters to stop changing. 
The final selected parameters are $a_\star=0.80$, $\lambda_\star = 0.75$, $p_\star=1.50$ and $q_\star=1.00$, which are used in our simulations. \textcolor{black}{We found that a different search order results in a different set of parameters. However, our simulations have shown that the final processing gain is virtually the same regardless of the order used.}

\subsection{Denoising results and analysis}

\begin{figure}[htb]
    \centering
     \includegraphics[width=\linewidth,trim={0 1cm 0 0}]{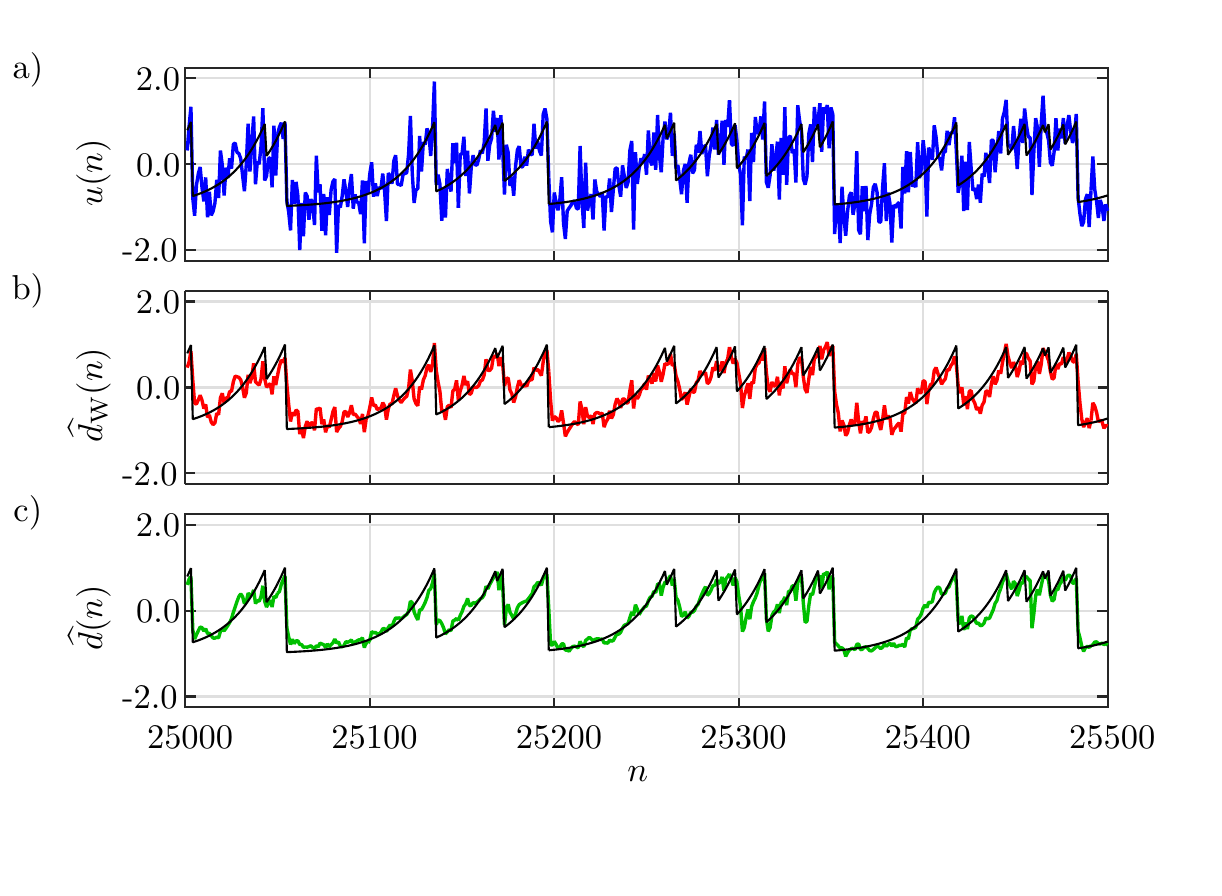}
    \caption{Noise reduction for $\alpha = 0.9$ and $\text{SNR}_\text{in}=$ 2.0~dB. Black lines represent the desired signal $d(n)$. a) Input signal $u(n)$; b) WF estimated signal $\widehat{d}_\text{W}(n)$; c) ESN estimated signal $\widehat{d}(n)$.}
    \label{fig:results2}
\end{figure}

Figure \ref{fig:results2} shows examples of the estimated signals using ESN and a WF \cite{HaykinAdaptiveFilterTheory} for $\text{SNR}_\text{in}=$ 2.0~dB and skew tent map parameter $\alpha=0.9$. We denote the signal estimated by the ESN $\widehat{d}(n)$ and the signal estimated by the WF by $\widehat{d}_\text{W}(n)$. In this case, $\text{SNR}_\text{out}= 7.7$~dB for the ESN  and $\text{SNR}_\text{out}= 5.1$~dB for the WF, showing that the ESN outperforms the WF in this scenario.

As can be seen in \eqref{eq:skewTentMapPSD} and Figure \ref{fig:chaoticSignalsFigure}(c), the PSD of the chaotic signal becomes whiter as $|\alpha|$ approaches zero. Thus, in terms of autocorrelation it becomes similar to the corrupting noise, denoising it becomes more difficult, and one can expect the processing gain to decrease with $|\alpha|$.

\begin{figure}[htb]
    \centering
    \includegraphics[width=\linewidth,trim={0 0cm 0 0}]{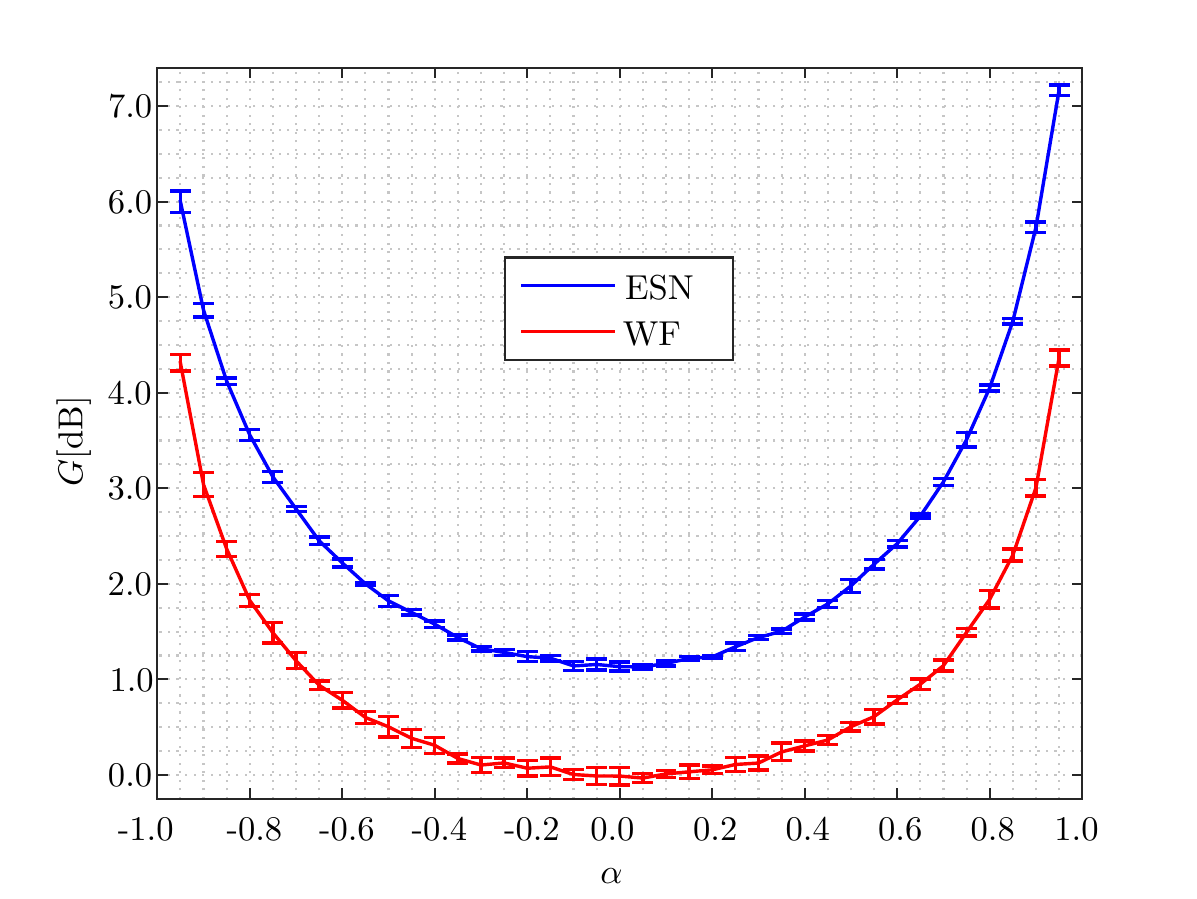}
    \caption{Processing gain in dB as a function of the skew tent map parameter $\alpha$ using an ESN and a WF.}
    \label{fig:results}
\end{figure}

Using the optimal parameters obtained in Section \ref{subsec:parameters}, the ESN was trained and tested for different values of $\alpha \in (-1, 1)$. Figure \ref{fig:results} presents the results obtained in terms of the mean processing gain after five training / test scenarios.  As a benchmark, a WF with 10 taps was used to perform the same denoising task. The error bars indicate the standard deviation calculated from the five repetitions. 

The presented results show that the ESN outperforms the WF for all values of $\alpha$. \textcolor{black}{This follows not only because ESN is a nonlinear technique, but also because it has proven to be well suited for handling time series \cite{Hunt,Xu2019}.} Furthermore,  as expected, the performance of both techniques degrades as $\alpha$ approaches zero. Note that for $\alpha=0$, the PSD of the chaotic signal is white, and the processing gain for the WF is zero\textcolor{black}{, unlike the processing gain for the ESN.}

\textcolor{black}{In addition to outperforming the WF for each $\alpha$ considered, the ESN processing gain increases as $|\alpha|$ increases, and at a faster rate}. The ESN processing gain is maximum for $\alpha=0.95$, where it reaches $7.17\pm 0.05$~dB.

 In papers such as \cite{Han2007,Lou2022} processing gains ranging from $6.0$ dB to $25.0$ dB have been obtained using other techniques but always considering continuous-time chaotic systems. As mentioned in Section \ref{sec:introd}, we consider the task of denoising discrete-time chaotic signals to be much more difficult.

\section{Conclusions}
\label{sec:conclusions}
In this paper, we have analyzed the use of an ESN to reduce noise in a chaotic signal. Unlike other works in the literature, we considered signals generated by discrete-time maps, whose lack of smoothness makes the task more challenging. 
As a chaotic signal generator, we considered the skew-tent map because the dependence of its PSD on its single parameter is known, which allows an analysis of the influence of spectral whiteness on the denoising performance.

We have shown that by selecting and tuning the network parameters, the ESN outperforms a WF, the optimal linear filtering technique, in all cases, especially when the PSD of the chaotic signal is white, resulting in zero gain for the WF. For increasing values of $|\alpha|$, the PSD of the chaotic signals becomes narrowband, and the obtained processing gain is higher.

\cblue{As future investigations, we intend to compare the performance of the current method with other nonlinear techniques, such as wavelet filtering or standard RNNs, and also to evaluate how the proposed noise reduction technique can be used to improve the performance of chaos-based communication systems in noisy channels.}
\section*{Acknowledgments} 

This work was partially supported by the
National Council for Scientific and Technological Development (CNPq-Brazil) (grant number 311039/2019-7) and by the Coordination for the Improvement of Higher Education Personnel (CAPES-Brazil) (grant number 001).

\begin{small}

\end{small}
\end{document}